\documentclass[12pt]{article}
\usepackage{epsfig}
\def\bea{\begin{eqnarray}}
\def\eea{\end{eqnarray}}
\def\beq{\begin{equation}}
\def\eeq{\end{equation}}

\def\as{\alpha_s}
\def\az{\alpha_0}
\relax
\def\MS{\hbox{$\overline{\rm MS}$}}  

\def\abs#1{\left|#1\right|}

\def\ie{{\it i.e.}}
\def\MS{\hbox{$\overline{\rm MS}$}}
\def    \hepph  #1 {{\tt hep-ph/#1}}
\def    \hepex  #1 {{\tt hep-ex/#1}}

\newcommand{\sect}[1]{\setcounter{equation}{0}\section{#1}}

\newsavebox\tmpfig

\begin{document}

\pagestyle{empty}

\begin{flushright}

{\tt hep-ph/0504115}\\IFUM-826/FT\\ GeF/TH/4-05\\
\end{flushright}

\begin{center}
\vspace*{0.5cm}
{\Large \bf Renormalization group approach\\
to Sudakov resummation
in prompt photon production} \\
\vspace*{1.5cm}
Paolo Bolzoni$^{a}$, Stefano Forte$^{a}$ and Giovanni Ridolfi$^{b}$\\
\vspace{0.6cm}  {\it
{}$^a$Dipartimento di Fisica, Universit\`a di Milano and
INFN, Sezione di Milano,\\
Via Celoria 16, I-20133 Milano, Italy\\ \medskip
{}$^b$Dipartimento di Fisica, Universit\`a di Genova and
INFN, Sezione di Genova,\\
Via Dodecaneso 33, I-16146 Genova, Italy}\\
\vspace*{1.5cm}

{\bf Abstract}
\end{center}

\noindent

We prove the all-order exponentiation of soft
logarithmic corrections to prompt photon production in hadronic
collisions, by generalizing an approach previously developed in the
context of
Drell-Yan production and deep-inelastic scattering. We show that
all large logs in the soft limit
can be expressed in terms
of two dimensionful variables, and we use the renormalization
group to resum them. 
The resummed  results that we obtain
are more general though less predictive than those
proposed by other groups, in that they can accommodate for violations
of Sudakov factorization.
\vspace*{1cm}

\vfill
\noindent

\begin{flushleft} April 2005 \end{flushleft}
\eject

\setcounter{page}{1} \pagestyle{plain}

\sect{Introduction}

The resummation of
logarithmically-enhanced terms in the perturbative expansion of QCD
cross sections near the boundary of phase space 
(Sudakov~\cite{sudakov} resummation, henceforth)
has been proved to next-to-leading
logarithmic order for deep-inelastic scattering in the $x\to1$ limit
and for Drell-Yan production near threshold~\cite{cnt,sterman}. 
In fact, resummation formulae can be
derived for a wider class of processes and to all logarithmic orders,
by assuming the validity of a suitable two-scale 
factorization~\cite{contopa}. The two scales correspond to   
the hard scale which guarantees the
perturbative nature of the process, and another dimensionful variable,
whose large ratio to the first perturbative scale is the quantity
whose log must be resummed to
all orders. This two-scale factorization, originally
derived for semi-inclusive two-jet production in $e^+e^-$
annihilation~\cite{cs},  can be proved for deep-inelastic scattering and
Drell-Yan in the  $\phi^3$ theory in
six space-time dimensions~\cite{fact}, but its generalization to
QCD processes is nontrivial. 

In ref.~\cite{fr} a new approach to Sudakov resummation was developed,
and applied to the Drell-Yan and deep-inelastic scattering processes.
This approach has the
advantage of being valid to all
logarithmic orders, and self-contained,  in that it does not require any
factorization beyond the standard factorization of collinear
singularities. It relies on an essentially kinematical analysis of the
phase space for the given process in the soft limit, which is used
to establish the result that the dependence on the resummation variable
only appears through a given fixed dimensionful combination. This
provides a second dimensionful variable, along with
the hard scale of the process, which can
be resummed using standard renormalization group techniques. Beyond
the leading log level, the
resummed result found within this approach turns out to be somewhat
less predictive than the result
of refs.~\cite{cnt,fact}: in those references the
next$^k$-to-leading log resummed result is fully determined by a fixed
next$^k$-to-leading order computation, whereas a higher fixed order
computation is needed to determine all coefficients in the resummed
formula of ref.~\cite{fr}. The more predictive result is recovered
within this approach if
the dependence of the perturbative coefficients on the two 
dimensionful variables factorizes, i.e. if the 
two-scale factorization mentioned above holds.

In this paper we discuss the generalization of the approach of
ref.~\cite{fr} to the resummation of  the inclusive
transverse momentum spectrum of prompt photons 
produced in hadronic
collisions in the region where the transverse momentum is close to its
maximal value. Prompt photon production is 
a less inclusive process than Drell-Yan or deep-inelastic scattering,
and it is especially interesting from the
point of view of the approach of ref.~\cite{fr}, because 
the large logs which must be resummed  turn out to depend on
{\it two} 
independent dimensionful variables, on top of the hard scale of the
process: 
hence, prompt photon production is
characterized by three scales.  The possibility that the general
factorization ref.~\cite{fact} might extend to prompt photon
production was discussed in ref.~\cite{dps}, based on previous
generalizations~\cite{bido} of factorization, and used to
derive the corresponding resummed results. Resummation formulae for
prompt photon production in the approach of ref.~\cite{cnt} were also
proposed in ref.~\cite{ctdp}, and some arguments which might support
such resummation were presented in ref.~\cite{ctb}. 
Our treatment will provide a
full proof of resummation to all logarithmic orders. Our resummation
formula does not require the factorization proposed in
refs.~\cite{dps,ctdp}, and it is accordingly less predictive. 
Because of
the presence of two scales, it is also weaker than the
result of ref.~\cite{fr} for  DIS and Drell-Yan production. Increasingly more predictive
results are recovered if increasingly restrictive forms of factorization
hold. 

\sect{Kinematics of prompt photon production in the soft limit}

We consider the process
\beq
\label{HHg}
H_1(P_1)+H_2(P_2)\rightarrow \gamma(p_\gamma)+X,
\eeq
specifically the differential cross section
$p_\perp^3\frac{d\sigma}{dp_\perp}$,
where $p_\perp$ is the transverse momentum of the photon with respect to
the direction of the colliding hadrons $H_1$ and $H_2$, and
\beq
\label{x1}
x=\frac{ 4p^2_\perp}{S};\qquad
S=(P_1+P_2)^2;\qquad
0\leq x_\perp\leq 1.
\eeq
The factorized expression for this cross section 
in perturbative QCD is
\beq
\label{crosssec}
p_\perp^3\frac{d\sigma}{dp_\perp}(x,p_\perp^2)
=\sum_{a,b}\int_0^1dx_1\,dx_2\,dz\,x_1F_{a}^{H_1}(x_1,\mu^2)\,
x_2F_{b}^{H_2}(x_2,\mu^2)\,
C_{ab}\left(z,\frac{Q^2}{\mu^2},\as(\mu^2)\right)\delta
\left(x-z x_1x_2\right),
\eeq
where $F_a^{H_1}(x_1,\mu^2)$, $F_b^{H_2}(x_2,\mu^2)$
are the distribution functions of partons $a,b$ in the colliding
hadrons,
we have defined
\bea
\label{qdef}
Q^2&=&4p_\perp^2,   \\
z&=&\frac{Q^2}{s},  \qquad 0\leq z\leq 1,
\label{xdef}
\eea
where $s$ is the center-of-mass energy of the partonic
process,
and the coefficient function\phantom{a} $C_{ab}(z,\frac{Q^2}{\mu^2},\as(\mu^2))$
is defined in terms of the partonic cross section for the process
where partons $a$, $b$ are incoming as
\beq
C_{ab}\left(z,\frac{Q^2}{\mu^2},\as(\mu^2)\right)
=p_\perp^3\frac{d\hat\sigma_{ab}}{dp_\perp}.
\eeq

We will study the cross section eq.~(\ref{crosssec}) in the limit in
which
$z\to1$, i.e., the transverse momentum of the photon is close to its
maximal value, thereby suppressing the phase space for further parton
radiation. 
The convolution in eq.~(\ref{crosssec}) 
is turned into an ordinary product by Mellin transformation:
\beq
\sigma(N,Q^2)
=\sum_{a,b}F_{a}^{H_1}(N+1,\mu^2)\,F_{b}^{H_2}(N+1,\mu^2)\,
C_{ab}(N,Q^2/\mu^2,\as(\mu^2)),
\label{cdef}
\eeq
where
\beq
\label{mellin}
\sigma(N,Q^2)=
\int_0^1d x\,x^{N-1}\,
p_\perp^3\frac{d\sigma}{dp_\perp}(x,p_\perp^2)
\eeq
and similarly for $C_{ab}$ and $F^{H_j}_i$.

Whereas the cross section $\sigma(N,Q^2)$ is clearly
$\mu^2$-independent, this is not the case for each  
contribution to it from individual parton
subprocesses. However, the $\mu^2$ dependence of each contribution to
the sum over $a,\>b$ in eq.~(\ref{crosssec})  is proportional to the
off-diagonal anomalous dimensions $\gamma_{qg}$ and
$\gamma_{gq}$. In the large $N$ limit, these are suppressed by a power
of $\frac{1}{N}$ in comparison to $\gamma_{gg}$ and
$\gamma_{qq}$, or, equivalently, the corresponding splitting functions
are suppressed by a factor of $1-x$ in the large $x$ limit.
Hence, in the large $N$ limit each parton subprocess
can be treated independently, specifically, each $C_{ab}$ is
separately renormalization-group invariant.  Because 
we are interested in the behaviour of $C_{ab}(N,Q^2/\mu^2,\as(\mu^2))$
in the limit $N\to \infty$, we shall henceforth omit the parton
indices $a,\>b$ and assume that each subprocess is being treated independently.
Furthermore, on top of eqs.~(\ref{crosssec},~\ref{cdef})
the physical process eq.~(\ref{HHg}) receives another
factorized contribution, in which the final-state photon is produced by
fragmentation of a primary parton produced in the partonic sub-process. 
However, the
cross section for this process is also suppressed by a factor of
$\frac{1}{N}$ in the large $N$ limit. This is due to the fact that  the
fragmentation 
function carries this suppression, for the same reason why the anomalous
dimensions 
$\gamma_{qg}$ and $\gamma_{gq}$ are suppressed. Therefore, we will
disregard the fragmentation
contribution.

Because resummation takes the form of an exponentiation, it is useful 
to introduce the  log derivative of
the cross section $\sigma$, \ie, the so-called physical anomalous
dimension defined as
\beq
\label{physad}
Q^2\frac{\partial \sigma(N,Q^2)}{\partial Q^2}=
\gamma(N,\as(Q^2))\,\sigma(N,Q^2).
\eeq
The physical anomalous dimension $\gamma$ eq.~(\ref{physad}) is
independent of factorization scale, and it is related
to the standard anomalous dimension $\gamma^{AP}$, defined by
\beq
\label{evolF}
\mu^2\frac{\partial F(N,\mu^2)}{\partial \mu^2}= 
\gamma^{\rm AP}(N,\as(\mu^2)) F(N,\mu^2),
\eeq
according to
\beq
\label{generic}
\gamma(N,\as(Q^2))
=\frac{\partial\ln C(N,Q^2/\mu^2,\as(\mu^2))}{\partial\ln Q^2}
=\gamma^{\rm AP}(N,\as(Q^2))
+\frac{\partial\ln C(N,1,\as(Q^2))}{\partial\ln Q^2},
\eeq
where we have schematically denoted by $F(N,\mu^2)$ the product
$F_a^{H_1}(N,\mu^2)\,F_b^{H_2}(N,\mu^2)$ that appears
in the cross section for the given parton subprocess, eq.~(\ref{cdef}).
In terms of the physical anomalous dimension, the cross section
can be
written as 
\beq
\label{pert2}
\sigma(N,Q^2)=K(N;Q_0^2,Q^2)\,\sigma(N,Q_0^2)
=\exp\left[E(N;Q_0^2,Q^2)\right]\sigma(N,Q_0^2),
\eeq
where
\bea
\label{esplit} 
E(N;Q_0^2,Q^2)
&=&\int_{Q_0^2}^{Q^2}\frac{dk^2}{k^2}\gamma(N,\as(k^2))\label{edef}\\
&=&\int_{Q_0^2}^{Q^2}
\frac{dk^2}{k^2}\gamma^{\rm AP}(N,\as(k^2))
+\ln C(N,1,\as(Q^2))-\ln C(N,1,\as(Q_0^2)).
\nonumber
\eea

In the large-$x$ limit,
the order-$n$ coefficient of the
perturbative expansion is dominated by terms proportional to
$\left[\frac{\ln^k(1-x)}{1-x}\right]_+$,
with $k\leq 2n-1$, that must be resummed to all orders.
Upon Mellin transformation, these lead to contributions
proportional to powers of $\ln\frac{1}{N}$. In the sequel, we will
consider the resummation of these contributions to all logarithmic
orders, and disregard all contributions to the cross section which
are suppressed by powers of $1-x$, i.e., upon Mellin transformation,
by powers of $\frac{1}{N}$.

The resummation is performed in two steps, in analogy
to ref.~\cite{fr}. First, we show that the origin of the large logs is
essentially kinematical: we identify the configurations which
contribute in the soft limit, we show by explicit computation that
large Sudakov 
logs are produced by the phase-space for real emission
with the required kinematics as logs of two dimensionful variables,
and we show that this conclusion is unaffected by virtual corrections.
Second, we resum the logs of these variables using the
renormalization group.

The $l$-th order correction to the leading $O(\alpha_s)$ partonic
process receives contribution from 
the emission of up to $l+1$ massless partons with momenta
$k_1,\ldots,k_{l+1}$, and
\beq
\label{cons}
p_1+p_2=p_\gamma+k_1+\ldots k_{l+1}.
\eeq
In the partonic center-of-mass frame, where
\beq
p_1=\frac{\sqrt{s}}{2}(1,0,0,1)\qquad\qquad
p_2=\frac{\sqrt{s}}{2}(1,0,0,-1)
\eeq
\beq
\label{parfot}
p_\gamma=(p_\perp\cosh\eta_\gamma,\vec{p}_\perp,p_\perp\sinh\eta_\gamma),
\eeq
we have
\beq
\label{cons2}
(p_1+p_2-p_\gamma)^2=\frac{Q^2}{x}(1-\sqrt{x}\cosh\eta_\gamma)
=\sum_{i,j=1}^{l+1} k_i^0k_j^0(1-\cos\theta_{ij})\geq 0,
\eeq
where 
$\theta_{ij}$ is the angle between $\vec{k}_i$ and
$\vec{k}_j$. Hence,
\beq
1\leq \cosh\eta_\gamma\leq\frac{1}{\sqrt x}.
\label{etagammalim}
\eeq
Therefore,
\beq
\label{soft}\sum_{i,j=1}^{l+1} k_i^0k_j^0(1-\cos\theta_{ij})
=\frac{Q^2}{2}(1-x)+O\left[(1-x)^2\right].
\eeq

Equation~(\ref{soft}) implies that in the soft limit the sum of scalar
products of momenta $k_i$ of emitted partons eq.~(\ref{cons2}) must vanish.
However, 
contrary to the case of deep-inelastic scattering or Drell-Yan, not
all momenta $k_i$ of the emitted partons can be soft as $x\to 1$,
because the 3-momentum of the photon must be balanced. 
Assume thus that  momenta
$k_i,i=1,\ldots,n; n<l+1$ are soft in the $x\to 1$ limit, while
momenta $k_i,i>n$ are non-soft. For the sake of simplicity, 
we relabel non-soft momenta as
\beq
k'_j =k_{n+j};\qquad 1\leq j\leq m+1;\qquad m=l-n.
\eeq
The generic kinematic configuration in the
$x=1$ limit is then
\bea
&&k_i=0\qquad 1\leq i\leq n
\nonumber\\
&&\theta_{ij}=0;\qquad \sum_{j=1}^{m+1} k^{'0}_j=p_\perp\qquad
\qquad 1\leq i,j\leq m+1.
\eea
for all $n$ between $1$ and $l$, namely, the configuration where at
least one momentum is not soft, and the remaining momenta are either
collinear to it, or soft.

With this labeling of the momenta, the phase space
can be written as
(see the Appendix of ref.~\cite{fr})
\bea
\label{genid}
&&d\phi_{n+m+2}(p_1+p_2;p_\gamma,k_1,\ldots,k_n,k'_1,\ldots,k'_{m+1})
\\
&=&\int_0^s
\frac{dq^2}{2\pi}d\phi_{n+1}(p_1+p_2;q,k_1,\ldots,k_n)
\int_0^{q^2}\frac{d{k'}^2}{2\pi}\,
d\phi_{m+1}(k';k'_1,\dots,k'_{m+1})\,d\phi_2(q;p_\gamma,k').
\nonumber
\eea
We shall now compute the phase space in the $x\to1$ limit.
Consider first the two-body phase space $d\phi_2$ in eq.~(\ref{genid}).
In the rest frame of $q$ we have
\bea
d\phi_2(q;p_\gamma,k')
&=&\frac{d^{d-1}k'}{(2\pi)^{d-1}2k^{'0}}\,
\frac{d^{d-1}p_\gamma}{(2\pi)^{d-1}2p_\gamma^0}\,
(2\pi)^d\delta^{(d)}(q-k'-p_\gamma)
\nonumber\\
&=&\frac{(4\pi)^\epsilon}{8\pi\Gamma(1-\epsilon)}
\frac{P^{1-2\epsilon}}{\sqrt{q^2}}\,\sin^{-2\epsilon}\theta_\gamma\,
d\abs{\vec p_\gamma}\,d\cos\theta_\gamma\,
\delta(|\vec{p}_\gamma|-P)\label{psfd},
\eea
where $d=4-2\epsilon$ and
\beq
P=\frac{\sqrt{q^2}}{2}\left(1-\frac{{k'}^2}{q^2}\right).
\label{pgammacm}
\eeq
Because momenta $k_i$, $i\leq n$ are soft,
up to terms suppressed by powers of $1-x$,
the rest frame of $q$
is the same as the center-of-mass frame of the incoming partons,
in which
\bea
&&|\vec{p}_\gamma|=p_\perp\,\cosh\eta_\gamma
\\
&&\cos\theta_\gamma=\tanh\eta_\gamma.
\eea
Hence,
\beq
d\phi_2(q;p_\gamma,k')
=\frac{(4\pi)^\epsilon}{8\pi\Gamma(1-\epsilon)}
\frac{(Q^2/4)^{-\epsilon}}{\sqrt{q^2}}
\,dp_\perp\,d\eta_\gamma\,
\delta\left(\cosh\eta_\gamma-\frac{2P}{\sqrt{Q^2}}\right).
\label{ps3}
\eeq
The conditions
\beq
\cosh\eta_\gamma=\frac{2P}{\sqrt{Q^2}}\geq 1;
\qquad {k'}^2\geq 0
\label{M2softlim}
\eeq
restrict the integration range to
\bea
Q^2\leq &q^2&\leq s\label{regint3}\\
0\leq &{k'}^2&\leq q^2-\sqrt{Q^2q^2}\label{regint4}.
\eea
It is now convenient to define new variables $u,v$ 
\bea
q^2=Q^2+u(s-Q^2)&=&Q^2\left[1+u(1-x)\right]+O((1-x)^2)
\label{Q2soft}
\\
{k'}^2=v(q^2-\sqrt{Q^2q^2})&=&Q^2 \frac{1}{2}uv(1-x)+O((1-x)^2)
\label{M2soft}
\\
0\leq u\leq 1&;&0\leq v\leq 1,
\eea
in terms of which
\beq
P=\frac{\sqrt{Q^2}}{2}\left[1+\frac{1}{2}u(1-v)(1-x)\right]+O\left[(1-x)^2\right].
\eeq
Thus, the two-body phase space eq.~(\ref{ps3})  up to subleading
terms is given by
\beq
d\phi_2(q;p_\gamma,k')
=\frac{(4\pi)^\epsilon}{8\pi\Gamma(1-\epsilon)}
\frac{(Q^2/4)^{-\epsilon}}{\sqrt{Q^2}}
\,dp_\perp\,d\eta_\gamma\,
\frac{\delta(\eta_\gamma-\eta_+)+\delta(\eta_\gamma-\eta_-)}
{\sqrt{u(1-v)(1-x)}},
\label{ps4}
\eeq
where
\beq
\eta_\pm=\ln\left(\frac{2P}{\sqrt{Q^2}}\pm
\sqrt{\frac{4P^2}{Q^2}-1}\right)=\pm\sqrt{u(1-v)(1-x)}.
\eeq

We now note  that 
the phase-space element $d\phi_{n+1}(p_1+p_2;q,k_1,\ldots,k_n)$
contains in the final state a system with large invariant
mass $q^2\geq Q^2$, plus a collection of $n$ soft partons: this same
configuration is encountered in the case of Drell-Yan pair production
in the limit $x_{DY}=q^2/s\to 1$, discussed in ref.~\cite{fr}.
Likewise, the phase space for the set of collinear partons
$d\phi_{m+1}(k';k'_1,\ldots,k'_{m+1})$ is the same
as the phase space for deep-inelastic scattering, where
the invariant mass of the initial state ${k'}^2$ vanishes
as $1-x$ (see eq.~(\ref{M2soft})). We may therefore use the results
obtained in ref.~\cite{fr}:\footnote{In the case of deep-inelastic
scattering, in ref.~\cite{fr} one of the outgoing parton momenta
($k'_{m+1}$, say) was identified with the momentum of the outgoing
quark and called $p'$, hence eq.~(\ref{discase}) is obtained from the
corresponding result in ref.~\cite{fr} by the replacement $p'\to
k'_{m+1}$.}
\bea
d\phi_{n+1}(p_1+p_2;q,k_1,\dots,k_n)&=&
2\pi\left[\frac{N(\epsilon)}{2\pi}\right]^{n}(q^2)^{-n(1-\epsilon)}
(s-q^2)^{2n-1-2n\epsilon}d\Omega^{(n)}(\epsilon)
\label{dycase}
\\
d\phi_{m+1}(k';k'_1,\ldots,k'_{m+1})&=&
2\pi\left[\frac{N(\epsilon)}{2\pi}\right]^m({k'}^2)^{m-1-m\epsilon}
d\Omega^{'(m)}(\epsilon),
\label{discase}
\eea
where $N(\epsilon)=1/(2(4\pi)^{2-2\epsilon})$ and
\bea
&&d\Omega^{(n)}(\epsilon)=d\Omega_1\dots
d\Omega_n\int_0^1dz_{n}z_{n}^{(n-2)+(n-1)(1-2\epsilon)}
(1-z_{n})^{1-2\epsilon}\dots
\int_0^1dz_2z_2^{1-2\epsilon}(1-z_2)^{1-2\epsilon}
\\
&&d\Omega^{'(m)}(\epsilon)=d\Omega'_1\dots
d\Omega'_m\int_0^1dz'_mz_m^{'(m-2)-(m-1)\epsilon}
(1-z'_m)^{1-2\epsilon}\dots
\int_0^1dz'_2z_2^{'-\epsilon}(1-z'_2)^{1-2\epsilon}.
\eea
The definition of the variables $z_i,z'_i$ is irrelevant
here and can be found in ref.~\cite{fr}. 

Equations~(\ref{Q2soft},\ref{M2soft}) imply
that the phase space depends on $(1-x)^{-\epsilon}$ through the two
variables
\bea
&&{k'}^2\propto Q^2 (1-x)\\
&&\frac{(s-q^2)^2}{q^2}\propto Q^2 (1-x)^2,
\eea
where the coefficients of
proportionality are dimensionless and $x$-independent.
By explicitly combining the two-body phase space
eq.~(\ref{ps4}) and the phase spaces for soft radiation
eq.~(\ref{dycase})
and for collinear radiation eq.~(\ref{discase}) we get
\bea
&&d\phi_{n+m+2}(p_1+p_2;p_\gamma,k_1,\ldots,k_n,k'_1,\ldots,k'_{m+1})
=(Q^2)^{n+m-(n+m+1)\epsilon}
\frac{dp_\perp}{p_\perp}\,
\frac{(1-x)^{2n+m-(2n+m)\epsilon}}{\sqrt{1-x}}
\nonumber\\
&&\qquad
2^{-m+m\epsilon}\,
\frac{(16\pi)^{-1+\epsilon}}{\Gamma(1-\epsilon)}
\left[\frac{N(\epsilon)}{2\pi}\right]^{n+m}\,d\eta_\gamma\,
d\Omega^{(n)}(\epsilon)\,d\Omega^{'(m)}(\epsilon)
\nonumber\\
&&\qquad
\int_0^1du\,
\frac{u^{m-m\epsilon}(1-u)^{2n-1-2n\epsilon}}{\sqrt{u}}
\int_0^1dv\,
\frac{v^{m-1-m\epsilon}}{\sqrt{1-v}}\,
[\delta(\eta_\gamma-\eta_+)+\delta(\eta_\gamma-\eta_-)].
\label{dphi}
\eea
In the limiting cases $n=0$ and $m=0$ we have
\bea
d\phi_1(p_1+p_2;q)&=&2\pi\delta(s-q^2)=\frac{2\pi}{Q^2(1-x)}\,\delta(1-u)
\label{dycase0}
\\
d\phi_1(k';p')&=&2\pi\delta({k'}^2)=\frac{4\pi}{Q^2u(1-x)}\,\delta(v);
\label{discase0}
\eea
the corresponding expressions for the phase space are therefore obtained
by simply replacing
\beq
(1-u)^{-1}\,d\Omega^{(n)}(\epsilon)\to \delta(1-u);
\qquad
v^{-1}\,d\Omega^{'(m)}(\epsilon)\to \delta(v)
\eeq
in eq.~(\ref{dphi}) for $n=0$, $m=0$ respectively.

The logarithmic dependence of the four-dimensional cross section on
$1-x$ is due to interference between powers of $(1-x)^{-\epsilon}$ and
$\frac{1}{\epsilon}$ poles in the $d$-dimensional
cross section. Hence, we must classify the dependence of the cross
section on powers of $(1-x)^{-\epsilon}$. We have established that 
in the phase space 
each real emission 
produces a factor of
$\left[Q^2(1-x)^2\right]^{1-\epsilon}$ if the
emission is soft and a factor of
$\left[Q^2(1-x)\right]^{1-\epsilon}$  if the emission is collinear. 
The squared amplitude  
can only depend on  $(1-x)^{-\epsilon}$ because
of loop integrations. This dependence for a generic proper 
Feynman diagram $G$
will in general appear~\cite{fr} through a
coefficient 
\beq\label{loops} A_G(P_E) = \left[D_G(P_E)\right]^{dL/2-I},\eeq
where $L$ and $I$ are respectively the number of loops and internal
lines in $G$, and $D_G(P_E)$ is a linear combination of all scalar
products $P_E$ of external momenta. In the soft limit it is easy to see, by
manipulations analogous to eq.~(\ref{cons2}), that all scalar products
which vanish as $x\to 1$ are either proportional to $Q^2(1-x)$ or to
$Q^2(1-x)^2$. Equation~(\ref{loops}) then implies that each loop
integration can carry at most 
a factor of $\left[Q^2(1-x)^2\right]^{-\epsilon}$ 
or $\left[Q^2(1-x)\right]^{-\epsilon}$.

This then proves that the 
perturbative expansion of the bare coefficient function takes the
form
\bea\label{cofun}
C^{(0)}(x,Q^2,\az,\epsilon)&=&
\alpha\az(Q^2)^{-\epsilon}
\sum_{l=0}^{\infty}\az^l\,C_l^{(0)}(x,Q^2,\epsilon)
\\
C_l^{(0)}(x,Q^2,\az,\epsilon)&=&
\frac{(Q^2)^{-l\epsilon}}{\Gamma(1/2)\sqrt{1-x}}
\sum_{k=0}^l
\sum_{k'=0}^{l-k}
C_{lkk'}^{(0)}(\epsilon)
(1-x)^{-2k\epsilon-k'\epsilon}\label{gencoefffunc},
\eea
where the coefficients $C_{lkk'}^{(0)}$ have poles in $\epsilon=0$
up to order $2l$, and the factor $1/\Gamma (1/2)$ was
introduced for later convenience.
Terms with $k+k'<l$ at order $\as^l$ are present in general because of
loops.
 The Mellin transform of eq.~(\ref{cofun}) can be performed using 
\beq
\int_0^1dxx^{N-1}(1-x)^{-1/2-2k\epsilon-k'\epsilon}
=\frac{\Gamma(1/2)}{\sqrt{N}}N^{2k\epsilon}N^{k'\epsilon}+O\left(\frac{1}{N}\right),
\eeq
with the result
\bea
&&C^{(0)}(N,Q^2,\az,\epsilon)
\label{C0}
\\
&&=\frac{\alpha\az(Q^2)^{-\epsilon}}{\sqrt{N}}\sum_{l=0}^\infty
\sum_{k=0}^l
\sum_{k'=0}^{l-k}
C_{lkk'}^{(0)}(\epsilon)
\left[\left(\frac{Q^2}{N^2}\right)^{-\epsilon}\az\right]^k
\left[\left(\frac{Q^2}{N}\right)^{-\epsilon}\az\right]^{k'}
\left[(Q^2)^{-\epsilon}\az\right]^{l-k-k'}+O\left(\frac{1}{N}\right).
\nonumber
\eea

\sect{Resummation from renormalization group improvement}

Equation~(\ref{C0}) shows that indeed as $N\to\infty$, up to $\frac{1}{N}$
corrections,
the  coefficient function depends on
$N$ through the two dimensionful variables $\frac{Q^2}{N^2}$ and
$\frac{Q^2}{N}$.   The argument henceforth is a rerun of that of
ref.~\cite{fr}, in this somewhat more general situation.
The argument is based on the observation that, because of collinear
factorization, 
the physical anomalous dimension
\beq
\gamma(N,\as(Q^2))=
Q^2\frac{\partial}{\partial Q^2 }\ln
C(N,Q^2/\mu^2,\as(\mu^2))
\label{paddef}
\eeq
is renormalization-group invariant and finite when expressed in terms
of the renormalized coupling $\as(\mu^2)$, related to $\alpha_0$ by
\bea
\az(\mu^2,\as(\mu^2))=\mu^{2\epsilon}\as(\mu^2)\,
Z^{(\as)}(\as(\mu^2),\epsilon),
\label{renas}
\eea
where $Z^{(\as)}(\as(\mu^2),\epsilon)$ is a power series in $\as(\mu^2)$.
Because $\alpha_0$ is manifestly independent of $\mu^2$,
eq.~(\ref{renas}) implies that the dimensionless combination
$(Q^2)^{-\epsilon}\az(\as(\mu^2),\mu^2)$ can depend on $Q^2$ only
through $\alpha_s(Q^2)$:
\beq
(Q^2)^{-\epsilon}\az(\as(\mu^2),\mu^2)
=\as(Q^2)\,Z^{(\as)}(\as(Q^2),\epsilon).
\label{rencou}
\eeq

Using eq.~(\ref{rencou}) in eq.~(\ref{C0}), the coefficient
function and consequently the physical anomalous dimension are seen to
be given by a 
power series in $\as(Q^2)$, $\as(Q^2/N)$ and $\as(Q^2/N^2)$:
\beq
\gamma(N,\as(Q^2),\epsilon)
=\sum_{m=0}^\infty\sum_{n=0}^\infty\sum_{p=0}^\infty
\gamma^R_{mnp}(\epsilon)\,
\as^m(Q^2)\,\as^n(Q^2/N^2)\,\as^p(Q^2/N).
\label{gamma}
\eeq
Even though the anomalous dimension is finite as $\epsilon\to0$, the
individual terms in the expansion eq.~(\ref{gamma}) are not
separately finite. However, if we
separate $N$-dependent and $N$-independent
terms in eq.~(\ref{gamma}):
\beq
\gamma(N,\as(Q^2),\epsilon)
=\hat\gamma^{(c)}(\as(Q^2),\epsilon)+\hat\gamma^{(l)}(N,\as(Q^2),\epsilon),
\label{gammahat}
\eeq
we note that the two functions
\bea
\gamma^{(c)}(\as(Q^2),\epsilon)
&\equiv&\hat\gamma^{(c)}(\as(Q^2),\epsilon)+\hat\gamma^{(l)}(1,\as(Q^2),\epsilon)
\\
\gamma^{(l)}(N,\as(Q^2),\epsilon)
&\equiv&\hat\gamma^{(l)}(N,\as(Q^2),\epsilon)-\hat\gamma^{(l)}(1,\as(Q^2),\epsilon)
\eea
must be separately finite, because 
\beq
\gamma(N,\as(Q^2),\epsilon)=\gamma^{(c)}(\as(Q^2),\epsilon)+\gamma^{(l)}(N,\as(Q^2),\epsilon),
\label{gamdec}
\eeq
is finite for all $N$, and 
$\gamma^{(l)}$ vanishes for $N=1$. 

We can rewrite conveniently 
\beq
\gamma^{(l)}(N,\as(Q^2),\epsilon)
=\int_1^N \frac{dn}{n}\,g(\as(Q^2),\as(Q^2/n^2),\as(Q^2/n),\epsilon),
\eeq
where
\beq
g(\as(Q^2),\as(Q^2/n^2),\as(Q^2/n),\epsilon)=n\frac{\partial}{\partial n}
\hat\gamma^{(l)}(n,\as(Q^2),\epsilon).
\label{gdef}
\eeq
is a Taylor
series in its arguments whose coefficients remain finite as
$\epsilon\to0$. 
In four
dimension we have thus
\bea
\gamma(N,\as(Q^2))&=&\gamma^{(l)}(N,\as(Q^2),0)
+\gamma^{(c)}(N,\as(Q^2),0)+O\left(\frac{1}{N}\right)
\nonumber\\
&=&\gamma^{(l)}(N,\as(Q^2),0)+O\left(N^0\right)\nonumber\\
&=&\int_1^N \frac{dn}{n}\,
g(\as(Q^2),\as(Q^2/n^2),\as(Q^2/n))+O\left(N^0\right),
\label{resformula}
\eea
where $g(\as(Q^2),\as(Q^2/n^2),\as(Q^2/n))\equiv 
\lim_{\epsilon\to0}g(\as(Q^2),\as(Q^2/n^2),\as(Q^2/n),\epsilon)$ is a generic Taylor series
of its arguments.

Renormalization group invariance thus implies that 
the physical anomalous dimension $\gamma$ eq.~(\ref{paddef})
depends on its three arguments $Q^2$, $Q^2/N$ and $Q^2/N^2$ only
through $\alpha_s$. Clearly, any function of $Q^2$ and $N$ can be
expressed as a function of $\as(Q^2)$ and $\as(Q^2/N)$  or
$\as(Q^2/N^2)$. The nontrivial statement, which endows
eq.~(\ref{resformula}) with predictive power, is that the
log derivative of $\gamma$,
$g(\as(Q^2),\as(Q^2/n^2),\as(Q^2/n))$ eq.~(\ref{gdef}),
is analytic in its three arguments. This immediately implies that when
$\gamma$ is computed at (fixed) order $\alpha_s^k$, it is a polynomial
in $\ln\frac{1}{N}$ of $k$-th order at most.

In order to discuss the factorization properties of our result
we write the function $g$ as
\bea
\label{ggdef}
g(\as(Q^2),\as(Q^2/n^2),\as(Q^2/n))
&=&g_1(\as(Q^2),\as(Q^2/n))+g_2(\as(Q^2),\as(Q^2/n^2))
\nonumber\\
&&\qquad +g_3(\as(Q^2),\as(Q^2/n),\as(Q^2/n^2))
\nonumber\\
g_1(\as(Q^2),\as(Q^2/n))&=&
\sum_{m=0}^\infty\sum_{p=1}^\infty 
g_{m0p}\,\as^m(Q^2)\,\as^p(Q^2/n)
\\
g_2(\as(Q^2),\as(Q^2/n^2))&=&\sum_{m=0}^\infty\sum_{n=1}^\infty
g_{mn0}\, \as^m(Q^2)\,\as^n(Q^2/n^2)
\nonumber\\
g_3(\as(Q^2),\as(Q^2/n),\as(Q^2/n^2))
&=&\sum_{m=0}^\infty\sum_{n=1}^\infty\sum_{p=1}^\infty
g_{mnp}\,\as^m(Q^2)\,
\as^n(Q^2/n^2)\,\as^p(Q^2/n).
\nonumber
\eea
The dependence on the resummation variables is fully factorized if
the bare coefficient functions has the factorized structure
\beq
\label{fstrong}
C^{(0)}(N,Q^2,\az,\epsilon)=C^{(0,c)}(Q^2,\az,\epsilon)\,
C_1^{(0,l)}(Q^2/N,\az,\epsilon)\,
C_2^{(0,l)}(Q^2/N^2,\az,\epsilon).
\eeq
This is argued to be the case in the approach of refs.~\cite{dps,ctdp}.
If this happens, the resummed anomalous dimension is given
by eq.~(\ref{resformula}) with all $g_{mnp}=0$ except $g_{0n0},g_{00p}$: 
\beq \label{resformulanason} 
\gamma(N,\as(Q^2))
=\int_1^N\frac{dn}{n}\,g_1(0,\as(Q^2/n))
+\int_1^N\frac{dn}{n}\,g_2(0,\as(Q^2/n^2)).
\eeq
Note that because the coefficient function depends on the parton
subprocess (compare eq.~(\ref{crosssec})) the factorization eq.~(\ref{fstrong})
applies to the coefficient function corresponding to each subprocess,
and the decomposition eq.~(\ref{resformulanason}) to the physical anomalous dimension
computed from each of these  coefficient functions.

A weaker form of factorization  is obtained assuming that in the soft limit the
$N$-dependent and $N$-independent parts of the coefficient function
factorize: 
\beq
\label{fweak}
C^{(0)}(N,Q^2,\epsilon)=C^{(0,c)}(Q^2,\az,\epsilon)\,
C^{(0,l)}(Q^2/N^2,Q^2/N,\az,\epsilon).
\eeq
This condition turns out to be satisfied~\cite{fr} in Drell-Yan and
deep-inelastic scattering to order $\alpha_s^2$. It holds in QED to
all orders~\cite{weinberg}
as a consequence of the fact that each  emission in the soft limit
can be described by
a universal (eikonal) factors, independent of the underlying
diagram. This eikonal structure of Sudakov radiation has been argued
in refs.~\cite{cnt,ctdp} to apply also to QCD.
If the factorized form eq.~(\ref{fweak}) holds, the coefficients
 $g_{mnp}$ eq.~(\ref{ggdef})  vanish for all
$m\not= 0$, and  the physical anomalous
dimension takes the form  
\bea \label{resformulaint} 
\gamma(N,\as(Q^2))
&=&\int_1^N\frac{dn}{n}\,g_1(0,\as(Q^2/n))
+\int_1^N\frac{dn}{n}\,g_2(0,\as(Q^2/n^2))
\nonumber\\
&&+\int_1^N\frac{dn}{n}\,g_3(0,\as(Q^2/n^2),\as(Q^2/n)).
\eea

It is interesting to observe that in the approach of
refs.~\cite{dps,ctdp} for processes where more than one colour
structure contributes to the cross-section, the factorization
eq.~(\ref{fstrong}) of the coefficient function is argued to take place
separately for each colour structure. This means that in such case
the exponentiation takes place for each colour structure independently,
i.e. the resummed cross section for each parton subprocess is in turn
expressed as a sum of factorized terms of the form of
eq.~(\ref{fstrong}). This happens for instance in the case of heavy
quark production~\cite{ctdp,hq}. In prompt photon production
different colour structures appear for the gluon-gluon subprocess
which starts at next-to-next-to-leading order, hence their separated
exponentiation would be relevant
for next-to-next-to-leading log resummed results.

When several colour structures contribute to a given parton
subprocess, the coefficients of the perturbative expansion
eq.~(\ref{C0}) for that process take the form
\bea
C_{lkk'}^{(0)}(\epsilon)=C_{lkk'}^{(0)\bf{1}}(\epsilon)
+C_{lkk'}^{(0)\bf{8}}(\epsilon),
\label{colco}
\eea
(assuming for definiteness that a colour singlet and octet
contribution are present) so that the coefficient function can be
written as a sum $C^{(0)}=C^{(0)}_{\bf{1}}+C^{(0)}_{\bf{8}}$. 
The argument which leads from eq.~(\ref{C0}) to the resummed result
eq.~(\ref{resformula}) then implies that exponentiation 
takes place for each
colour structure independently if and only if 
$\gamma_{\bf{1}}\equiv\partial\ln C^{(0)}_{\bf{1}}/\partial\ln Q^2$ 
and 
$\gamma_{\bf{8}}\equiv\partial\ln C^{(0)}_{\bf{8}}/\partial\ln Q^2$ 
are separately finite. 

This, however, is clearly a more restrictive
assumption than that under which we have derived the result
eq.~(\ref{resformula}), namely that the full anomalous dimension
$\gamma$ is finite. 
It follows that exponentiation of each
colour structure must be a special case of our result. However, this
can only be true if the coefficients $g_{ijk}$ of the expansion
eq.~(\ref{ggdef}) of the physical anomalous dimension satisfy suitable
relations. In particular, at the leading log level, it is easy to see
that exponentiation of each colour structure is compatible with
exponentiation of their sum only if the leading order coefficients
are the same for the given colour
structures:
$g_{001}^{\bf{1}}=g_{001}^{\bf{8}}$ and
$g_{010}^{\bf{1}}=g_{010}^{\bf{8}}$.
This is indeed the case for heavy quark production (where $g_{001}=0$).

Note that, however, if the factorization holds for each colour
structure separately it will not apply to the sum of colour
structures. For instance, the weaker form of factorization
eq.~(\ref{fweak}) requires that
$C_{lkk'}^{(0)}(\epsilon)=F_{k+k'}(\epsilon)G_{l-k-k'}(\epsilon),$ but
 \beq
F^{\bf{1}}_{k+k'}(\epsilon)G^{\bf{1}}_{l-k-k'}(\epsilon)+F^{\bf{8}}_{k+k'}(\epsilon)G^{\bf{8}}_{l-k-k'}(\epsilon)\neq
F_{k+k'}(\epsilon)G_{l-k-k'}(\epsilon).
\eeq
Hence, our result eq.~(\ref{resformula}) for the sum of colour
structures is more general than the 
separate exponentiation of individual colour structures, but it leads
to results which have weaker factorization properties.

The resummation  coefficients $g_{mnp}$ can be determined by comparing 
the expansion of the resummed anomalous dimension $\gamma$ in powers of
$\as(Q^2)$ with a fixed-order calculation:
\beq
\gamma_{\rm FO}(N,\as)=\sum_{i=1}^{k_{\rm min}}\as^i\,\sum_{j=1}^i 
\gamma^i_j\,\ln^j\frac{1}{N}+O(\as^{k_{\rm min}+1})+O(N^0),
\label{fixedorder}
\eeq
where $\gamma_{\rm FO}(N,\as)$ is the physical anomalous dimension for
the same individual partonic subprocess (recall eq.~(\ref{crosssec})).
Clearly, if the more restrictive factorized forms eq.~(\ref{fstrong})
or eq.~(\ref{fweak}) hold, a smaller number of coefficients determines
the resummed result, and thus a lower fixed-order calculation is
sufficient to predict higher-order logarithmic terms than if the more
general eq.~(\ref{ggdef}) is used. Conversely, a higher fixed-order
calculation can be used to verify if the strong factorization
eq.~(\ref{fstrong}) holds as advocated in refs.~\cite{dps,ctdp}, or
whether one must use the less predictive but more general result
eq.~(\ref{ggdef}) that we have derived.

Once the resummed physical anomalous dimension has been determined,
the resummed cross section can be obtained from it using
eq.~(\ref{pert2}), with a factorization scheme choice which specifies
the way it is split into its two components
eq.~(\ref{generic}). Commonly used choices are the physical scheme
choice, in which $C=1$ so $\gamma=\gamma^{\rm AP}$, or the \MS\
scheme, in which the unresummed and resummed forms of the anomalous
dimensions $\gamma^{\rm AP}$ coincide. An explicit construction of the
relation between physical anomalous dimension and resummed cross
section, and the matching between resummed and unresummed results, can
be found in Section~6 of ref.~\cite{fr}.

\sect{The structure of resummed results}

We determine the predictive power of each resummed result by means of
the following strategy. First, we assume that the
coefficients $g_{mnp}$ needed for N$^{k-2}$LL resummation have already
been determined. Next, we identify the coefficients that are needed to
extend the accuracy to N$^{k-1}$LL, and we write a system of equations
that fix them in terms of the known coefficients, and of the
$\gamma^i_j$ of the fixed-order expansion.  The rank of this system of
equations determines the minimum order $k_{\rm min}$ in $\as$ of a fixed
order computation which is needed to fix the N$^{k-1}$LL resummation. This
means that at any higher
fixed order $f>k_{\rm min}$, the coefficients of all powers of
$\left(\ln\frac{1}{N}\right)^n$ with $k<n\le f$ are then predicted by
the resummed formula. 

The general structure of the anomalous dimension resummed to 
N$^{k-1}$LL accuracy is 
\bea
&&\gamma(N,\as(Q^2))=\sum_{p=1}^k\gamma_p(N,\as(Q^2))
\label{gammagen}
\\
&&\gamma_p(N,\as(Q^2))=
\label{gammas}
\sum_{i=0}^{p-1}\sum_{j=0}^{p-i}
g_{ijp-i-j}\,\as^i(Q^2)\,
\int_1^N\frac{dn}{n}\,\as^j(Q^2/n^2)\,\as^{p-i-j}(Q^2/n).
\eea
At the N$^{k-1}$LL order, in each term $\gamma_p(N,\as(Q^2))$ the
coupling constant $\as(Q^2/n^a)$ 
can be expanded in powers of $\as(Q^2)$
using the N$^{k-p}$LL solution of the renormalization group equation
\beq
\mu^2\frac{d\as}{d\mu^2}=-\beta_0\as^2-\beta_1\as^3+\ldots \,,
\label{beta}
\eeq
because subsequent orders would lead to N$^{k}$LL contributions to
$\gamma$. 
In particular, the
leading log expression
\beq
\label{llas}
\as(Q^2/n^a)=\frac{\as(Q^2)}{1+a\as(Q^2)\beta_0\ln\frac{1}{n}}
\eeq
is sufficient for the computation of $\gamma_k(N,\as(Q^2))$. With
$\as(Q^2/n^a)$ given by eq.~(\ref{llas}) one gets
\bea
&&\int_1^N\frac{dn}{n}\,
\as^i(Q^2/n^2)\,\as^j(Q^2/n)
=\sum_{m=0}^\infty C^{(i,j)}_m\,\beta_0^m\,
\as(Q^2)^{i+j+m}\,\ln^{m+1}\frac{1}{N}
\\
&&
C^{(i,j)}_m=\frac{(-1)^{m+1}}{m+1}\sum_{l=0}^m 2^l
\left(\begin{array}{c}l+i-1\\i-1\end{array}\right)
\left(\begin{array}{c}m-l+j-1\\j-1\end{array}\right)
\label{Cijm}
\eea
(note that 
\hbox{$\left(\begin{array}{c}n\\-1\end{array}\right)=1$} for $n=-1$
and $0$ otherwise), and therefore
\beq
\gamma_k(N,\as(Q^2))=
\sum_{i=0}^{k-1}\sum_{j=0}^{k-i}
g_{ijk-i-j}\,
\sum_{m=0}^\infty C^{(j,k-i-j)}_m
\,\beta_0^m\as(Q^2)^{k+m}\,\ln^{m+1}\frac{1}{N}.
\label{gammak}
\eeq

Let us consider first the leading log case, $k=1$. In this case, $g$
is linear in $\as$ and therefore eq.~(\ref{resformulanason}) always
holds, i.e. there is no distinction between factorized and
unfactorized results. The anomalous dimension has the form
\beq
\gamma_1(N,\as(Q^2))
=\sum_{m=0}^\infty \left(g_{001}\,+2^mg_{010}\right)\,
\frac{(-1)^{m+1}}{m+1}\,\beta_0^m
\as^{m+1}(Q^2)\,\ln^{m+1}\frac{1}{N}+\hbox{NLL}.
\label{gamma1}
\eeq
Comparing with the fixed order expansion, we find
\bea
&&m=0:\qquad \gamma^1_1=-(g_{001}+g_{010})
\label{g001}
\\
&&m=1:\qquad \gamma^2_2=\frac{\beta_0}{2}\,(g_{001}+2g_{010}).
\label{g010}
\eea
These two conditions determine $g_{001}$ and $g_{010}$:
leading-log
resummation requires computing $\gamma$ to order $\as^2$. Note that
for DIS and Drell-Yan an $O(\as)$ computation is instead
sufficient~\cite{fr}, because only one scale is present and thus only
one coefficient has to be determined.

Let us now examine higher logarithmic orders, by discussing the various
factorizations in turn. Assume first the validity of the 
most restrictive result eq.~(\ref{resformulanason}), where the
dependence on the three scales $Q^2$, $Q^2/N$ and $Q^2/N^2$ is fully
factorized. In such case,  the
anomalous dimension is just the sum of a function of $\as(Q^2/N)$ and a
function of $\as(Q^2/N^2)$, and 
all coefficients $g_{mnp}$ vanish except $g_{00p},\>g_{0p0}$. In this
case,
$N_k=2k$ coefficients
are required for the N$^{k-1}$LL resummation. 
According to the strategy outlined above, we now assume that the
coefficients $g_{00p},g_{0p0}$ with $p\leq k-1$, relevant for
N$^{k-2}$LL resummation, have already been determined. The two extra
coefficients $g_{00k},g_{0k0}$ appear in $\gamma_k(N,\as)$, whose
explicit form is
given by eq.~(\ref{gammak}) with only $g_{00k},g_{0k0}$ different
from zero:
\beq
\gamma_k(N,\as(Q^2))=\sum_{m=0}^\infty
(C^{(0,k)}_m\,g_{00k}+C^{(k,0)}_m\,g_{0k0})\beta_0^m\,
\as^{k+m}(Q^2)\ln^{m+1}\frac{1}{N}.
\label{catanik2}
\eeq
Other terms of order $\as^{k+m}\ln^{m+1}(1/N)$ are generated by
expanding the coupling
$\as(Q^2/n^a)$ in $\gamma_1,\ldots,\gamma_{k-1}$ up to N$^{k-1}$LL;
however, these are fully  determined by the 
known coefficients $g_{00i},g_{0i0},i\leq k-1$ (and by the 
coefficients of the beta function $\beta_0,\ldots,\beta_{k-1}$).
Equating terms of order $\as^{k+m}\ln^{m+1}(1/N)$
in the fixed-order and resummed expressions of $\gamma$ we get the set 
of equations 
\bea
&&m=0:\qquad \gamma^k_1=-(g_{00k}+g_{0k0})\,
                 +F^{(k)}_1(g_{00i},g_{0i0};1\leq i\leq k-1) 
\\
&&m=1:\qquad \gamma^{k+1}_2=(g_{00k}+2g_{0k0})\,\frac{k\beta_0}{2}
                 +F^{(k)}_2(g_{00i},g_{0i0};1\leq i\leq k-1) 
\\
&&\ldots ,
\nonumber
\eea
where $F^{(i)}_j$ are known functions of the coefficients $g$ which we
have assumed to be already known. 

Hence, the two terms $m=0,1$ provide
two independent conditions 
that fix $g_{00k}$ and $g_{0k0}$ in terms of $g_{00i},\>g_{0i0},\>i\leq k-1$.
The same procedure can be repeated for $p=k-1,k-2,\ldots,1$; at each step,
$g_{00p}$ and $g_{0p0}$ are computed as functions of 
$\gamma^p_1,\gamma^{p+1}_2$ and $g_{00i},\>g_{0i0},\>0\leq i\leq p-1$. 
We conclude that in the case of eq.~(\ref{resformulanason})
the coefficients $g_{00i},g_{0i0};0\leq i\leq k$, relevant for N$^{k-1}$LL
resummation, are obtained from the fixed-order expansion of $\gamma$
up to order $k_{\rm min}=k+1$ (corresponding to $m=1$ in
eq.~(\ref{catanik2})). This means that even though at each extra
logarithmic order two
new coefficients appear, a single extra fixed order in $\alpha_s$ is
sufficient to determine both of
them.

Consider  for example the
next-to-leading log resummation. In our approach, this requires
the calculation of the physical anomalous dimension up to order
$\as^3$. Explicitly,
\bea
\gamma_1(N,\as(Q^2))
&=&-(g_{001}+g_{010})\,\as(Q^2)\,\ln\frac{1}{N}
\nonumber\\
&&+\frac{\beta_0}{2}(g_{001}+2g_{010})\,\as^2(Q^2)\,\ln^2\frac{1}{N}
-\frac{\beta_1}{\beta_0} (g_{001}+g_{010})\,\as^2(Q^2)\,\ln\frac{1}{N}
\nonumber\\
&& 
-\frac{\beta_0^2}{3}(g_{001}+4g_{010})\,\as^3(Q^2)\,\ln^3\frac{1}{N}
+\frac{3\beta_1}{2}(g_{001}+2g_{010})\,\as^3(Q^2)\,\ln^2\frac{1}{N}
\\
\gamma_2(N,\as(Q^2))
&=&-(g_{002}+g_{020})\,\as^2(Q^2)\ln\frac{1}{N}
+\beta_0(g_{002}+2g_{020})\,\as^3(Q^2)\ln^2\frac{1}{N}.
\label{NLL}
\eea
The leading log coefficients are fixed by the linear and quadratic
$\ln\frac{1 }{N}$ terms  in $\gamma_1$, 
eqs.~(\ref{g001},\ref{g010}), and the NLL coefficients by the the linear and quadratic
$\ln\frac{1 }{N}$ terms  in $\gamma_2$:
\bea
&&
g_{002}+g_{020}=-\gamma^2_1-\frac{\beta_1}{\beta_0}\,(g_{001}+g_{010})
\label{g002}
\\
&&
g_{002}+2g_{020}
=\frac{1}{\beta_0}\gamma^3_2-\frac{3}{2}\frac{\beta_1}{\beta_0}\,
(g_{001}+2g_{010}).
\label{g020}
\eea
All other logarithmically enhanced  contributions to the order $\as^3$
anomalous dimension are predicted, except the $\as^3\ln\frac{1 }{N}$
which depends on the N$^2$LL coefficients. The same
pattern continues at higher orders.

It is interesting to observe that even assuming the fully factorized
form of the coefficient function eq.~(\ref{fstrong}), our
results are still less restrictive and thus less predictive than those of
refs.~\cite{dps,ctdp}, where one less fixed order is required to
determine the resummed result.  For instance, at NLL,  using the notation of
ref.~\cite{ctdp}, the  resummation coefficients are given by
\bea
g_{010}&=&-\frac{A_a^{(1)}+A_b^{(1)}-A_d^{(1)}}{\pi}
\label{primoconfa}\\\label{primoconfb}
g_{001}&=&-\frac{A_d^{(1)}}{\pi}\\
g_{020}&=&-\bigg[\frac{A_a^{(2)}+A_b^{(2)}-A_d^{(2)}}{\pi^2}
-\frac{\beta_0(A_a^{(1)}+A_b^{(1)}-A_d^{(1)})}{\pi}\ln2-2\gamma_{E}\beta_0
\frac{A_a^{(1)}+A_b^{(1)}-A_d^{(1)}}{\pi}\bigg]\\
g_{002}&=&-\bigg[\frac{A_d^{(2)}}{\pi^2}
-\frac{\beta_0B_d^{(1)}}{2\pi}-\gamma_{E}\beta_0\frac{A_d^{(1)}}{\pi}\bigg],
\label{ultimoconfronto}
\eea
where 
$A_a^{(i)}$ is the coefficient of $\ln(1/N)$ in the Mellin transform
of the $P_{aa}$ Altarelli-Parisi splitting function
at order $\as^i$,
 $\gamma_E$ is the Euler constant, and $B^{(1)}_d$ is a
constant to be determined from the comparison with the fixed-order
calculation. In eq.~(\ref{primoconfa}-\ref{ultimoconfronto}) $a,\>b$ are the incoming
partons (on which $C$ implicitly depends), and $d$ is the
outgoing parton in the leading order process when
incoming partons $a$ and $b$ (which is uniquely determined by $a$ and
$b$).

Thus, in this approach $g_{0i0}$ is entirely determined in terms of the
$O(\alpha_s^i)$ coefficient of the $\ln(1/N)$ term in the 
anomalous dimension, and only $g_{00i}$ must be determined by
comparison to the fixed-order calculation: at the LL level, the
resummed result is obtained from the knowledge of $\gamma_{\rm FO}$ to
order $\as$,
at the NLL level to
order $\as^2$ and so on. This means that to LL, the coefficient
$\gamma_2^2$ eq.~(\ref{g010}) is in fact predicted by
eqs.~(\ref{primoconfa},\ref{primoconfb}) in terms of the coefficients
of the Altarelli-Parisi splitting functions. This prediction is borne
out by the explicit $O(\alpha_s^2)$ calculation of the prompt photon
production cross section~\cite{dgnlo}.
At the NLL level,
the coefficient
$\gamma^3_2$ of $\as^3\ln^2(1/N)$,  is predicted:
\bea
\gamma_2^3&=&-\beta_0\Bigg[\frac{2(A_a^{(2)}+A_b^{(2)})-A_d^{(2)}}{\pi^2}
-\beta_0(2\ln2+4\gamma_{E})\frac{A_a^{(1)}+A_b^{(1)}}{\pi}\\ &&\quad+\beta_0
(2\ln2+3\gamma_{E})\frac{A_d^{(1)}}{\pi} \nonumber
-\frac{\beta_0B_d^{(1)}}{2\pi}\Bigg]-\beta_{1}\frac{3}{2}
\left[\frac{2(A_a^{(1)}+A_b^{(1)})-A_d^{(1)}}{\pi}\right].
\eea
The correctness of this result could be tested by an order $\as^3$
calculation. If it were to fail, the more general resummation formula
with $g_{020}$ determined by eq.~(\ref{g020}) should be used, or one
of the  resummations which do not assume the factorization
eq.~(\ref{fstrong}). 

Assume now that the weaker factorization
eq.~(\ref{fweak}) applies.
In this case, only $g_{0nq};1\leq n+q\leq k$ are nonzero. This amounts to
keeping only the term $i=0$ in the general expression eq.~(\ref{gammas})
for $\gamma_p(N,\as(Q^2))$.
The total number of coefficients at N$^{k-1}$LL is
\beq
\label{weakno}
N_k=\sum_{p=1}^k (p+1)=\frac{k(k+3)}{2}.
\eeq
In order to improve the accuracy from N$^{k-2}$LL to N$^{k-1}$LL,
$k+1$ new coefficients are needed, namely $g_{0ik-i}$ for $i=0,\ldots,k$.
As in the previous case, we isolate the N$^{k-1}$LL 
terms that depend on the new coefficients.
All such terms are contained in $\gamma_k$, which in this case
is given by the term $i=0$ in eq.~(\ref{gammak}):
\beq
\gamma_k(N,\as(Q^2))=
\sum_{j=0}^k g_{0jk-j}
\sum_{m=0}^\infty C^{(j,k-j)}_m\,\beta_0^m\as(Q^2)^{k+m}\,\ln^{m+1}\frac{1}{N}.
\label{m}
\eeq
The first $k+1$ terms in the sum over $m$ provide the set of
conditions
\bea\label{condzero}
&&m=0:\qquad \gamma^k_1=\sum_{j=0}^k g_{0jk-j}C^{(j,k-j)}_0
+G^{(k)}_1(g_{0jk-j};0\leq j\leq k-1)
\\
&&m=1:\qquad \gamma^{k+1}_2=\sum_{j=0}^k g_{0jk-j}C^{(j,k-j)}_1\,\beta_0
+G^{(k)}_2(g_{0jk-j};0\leq j\leq k-1)\\
&&\ldots\nonumber
\\\label{condkappa}
&&m=k:\qquad \gamma^{2k}_{k+1}=\sum_{j=0}^k g_{0jk-j}C^{(j,k-j)}_k\,\beta_0^k
+G^{(k)}_{k+1}(g_{0jk-j};0\leq j\leq k-1),
\eea
where again $G^{(i)}_j$ are known functions of the coefficients $g$ which we
have assumed to be already known. 
Equations~(\ref{condzero}-\ref{condkappa}) are linearly independent
(see the Appendix for an explicit proof), and therefore 
determine $g_{0ik-i};0\leq i\leq k$
in terms of $g_{0jk-j};0\leq j\leq {k-1}$. It follows that a computation of
$\gamma$ up to order $k_{\rm min}=2k$ is necessary for the
determination of the $\frac{k(k+3)}{2}$ coefficients needed for
N$^{k-1}$LL resummation: even
though the number of coefficients which must be determined grows
quadratically according to eq.~(\ref{weakno}), the required order in
$\alpha_s$ of the computation which determines them grows only
linearly.

Note that in fact a similar argument can be applied to the
cases of deep-inelastic scattering
and Drell-Yan in the absence of factorization, discussed in ref.~\cite{fr}. 
In that reference
it was observed that $k(k+1)/2$ coefficients must be known in order
to determine
the N$^{k-1}$LL result, and thus a fixed N$^{k(k+1)/2-1}$LO computation is
sufficient to determine them. However, using the argument given here
it is easy to see that in fact a calculation
of $\gamma$ up to fixed order $k_{\rm min}=2k-1$ only is sufficient to
determine all these coefficients. 
Consider for example the case of deep-inelastic scattering.
To N$^{k-1}$LL one finds
\bea
&&\gamma(N,\as(Q^2))=\sum_{p=1}^k\gamma_p(N,\as(Q^2))
\\
&&\gamma_p(N,\as(Q^2))=\sum_{i=0}^{p-1}g_{ip-i}\,\as^i(Q^2)\,
\int_1^N\frac{dn}{n}\,\as^{p-i}(Q^2/n).
\eea
In order to determine the $k$ coefficients 
$g_{0k},\ldots,g_{k-1 1}$ that are
needed to improve the accuracy from  N$^{k-2}$LL to  N$^{k-1}$LL,
we proceed as in the previous case: the new coefficients appear only
in
\beq
\gamma_k(N,\as(Q^2))=\sum_{i=0}^{k-1}g_{ik-i}\,
\sum_{m=0}^\infty C^{(0,k-i)}_m\,\beta_0^m\,
\as(Q^2)^{k+m}\,\ln^{m+1}\frac{1}{N}.
\label{DISDY}
\eeq
Each term in  the sum over $m$ in eq.~(\ref{DISDY}) provides an independent
condition on the coefficients $g_{ij}$ 
(the linear independence of these conditions is straightforwardly
proved in the Appendix). Hence, in order to determine all
 $g_{ik-i};0\leq i\leq k-1$ it is sufficient to determine all terms up
to $m=k-1$ in $\gamma_k$, i.e.  compute $\gamma$ up to order $\as^{2k-1}$.
The same happens in the case of Drell-Yan, which is obtained by replacing 
$C_m^{(0,p-i)}$ with $C_m^{(p-i,0)}$ in eq.~(\ref{DISDY}).

Let us now consider the most general case, in which the 
coefficient function does not satisfy  any factorization property.
Then, to N$^{k-1}$LL, the anomalous dimension
eqs.~(\ref{gammagen},\ref{gammas}) depends on 
\beq
\label{nofnum}
N_k=\sum_{p=1}^k\frac{p(p+3)}{2}=\frac{k(k+1)(k+5)}{6}
\eeq
coefficients overall; out of these, the new ones
which must be determined in order to go
from N$^{k-2}$LL to N$^{k-1}$LL are  the $k(k+3)/2$ coefficients
\beq
\label{nklog}
g_{ijk-i-j};\qquad i=0,\ldots,k-1;\;j=0,\ldots,k-i.
\eeq
These new coefficients are
contained in $\gamma_k$, now given by its general expression
eq.~(\ref{gammak}), and each term with fixed $m$ in the
expansion of $\gamma_k$ eq.~(\ref{gammak}) provides a new condition
on these coefficient. However, 
these conditions are not linearly independent for any choice of
$m$: rather, the rank of the matrix which gives the linear combination
of coefficients eq.~(\ref{nklog}) to be determined turns out to be
$2k$ (see the Appendix). This means that the  N$^{k-1}$LL order
resummed result depends only on $2k$ independent 
linear combinations of the $k(k+3)/2$ coefficients
eq.~(\ref{nklog}). 
Because a term with fixed $m$ in $\gamma_k$ is
of order $\as^{k+m}$, this implies that a computation of the
anomalous dimension up to fixed order $k_{\rm min}=3k-1$ is sufficient
for the N$^{k-1}$LL resummation. Note that when going from N$^{k-1}$LL
to N$^{k}$LL and $\gamma_k$ is now determined at this higher order,
in general some new linear combinations of the
$k(k+3)/2$ coefficients eq.~(\ref{nklog}) will appear through terms
depending on $\beta_1$. Hence, some of the combinations of
coefficients that were left undetermined in the N$^{k-1}$LL
resummation will now become determined. However, this does not affect
the value $k_{\rm min}$ of the fixed-order accuracy needed to push the resummed
accuracy at one extra order. 
In conclusion, even in the absence of any factorization, despite the
fact that now the number of coefficients which must be determined grows
cubically according to eq.~(\ref{nofnum}), the required order in
$\alpha_s$ of the computation which determines them grows only
linearly. 
 
The number of coefficients $N_k$ that must be determined at each
logarithmic order, and the minimum fixed order which is necessary in
order to determine them are summarized in Table 1, according to
whether the coefficient function is fully factorized
[eq.~(\ref{resformulanason})], or has factorized $N$-dependent and
$N$-independent terms [eq.~(\ref{resformulaint})], or not factorized
at all [eq.~(\ref{resformula})]. In the approach of
refs.~\cite{dps,ctdp} the coefficient function is fully factorized,
and furthermore some resummation coefficients are related to universal
coefficients of Altarelli-Parisi splitting functions, so that
$k_{\rm min}=k$.
For completeness, we also list in the table  the results for DIS and
Drell-Yan, according to whether the coefficient function has
factorized $N$-dependent and
$N$-independent terms (as in refs.~\cite{cnt,sterman}) or no
factorization properties (as in ref.~\cite{fr}). Current fixed-order
results support factorization for Drell-Yan and DIS only to the lowest
nontrivial order $O(\as^2)$. For prompt-photon production, available results
do not allow to test factorization, and test relation of resummation
coefficients to Altarelli-Parisi coefficients only to lowest $O(\as)$.
\begin{table}
\begin{center}
\begin{tabular}{|c||c|c|c||c|c|}
\hline
     & \multicolumn{3}{|c||}{Prompt photon} &
     \multicolumn{2}{|c|}{DIS, DY} \\
\hline
      & eq.~(\ref{resformulanason})& eq.~(\ref{resformulaint})& eq.~(\ref{resformula})
& refs.~\cite{cnt,sterman} & ref.~\cite{fr}\\
\hline\hline
$N_k$ & $2k$ & $\frac{k(k+3)}{2}$  & $\frac{k(k+1)(k+5)}{6}$ & $k$ & $\frac{k(k+1)}{2}$\\
\hline
$k_{\rm min}$ & $k+1$ & $2k$ & $3k-1$ & $k$ & $2k-1$\\
\hline
\end{tabular}
\caption{\label{tab} Number of
coefficients $N_k$ and minimum order of the required perturbative calculation
$k_{\rm min}$ for different versions of the N$^{k-1}$LL resummation.}
\end{center}
\end{table}

\sect{Conclusion}
In this paper, we have presented a generalization to prompt photon
production of the approach to Sudakov resummation which was introduced
in ref.~\cite{fr} for deep-inelastic scattering and Drell-Yan production. The
advantage of this approach is that it does not rely on factorization
of the physical cross section, and in fact it simply follows from
general kinematic properties of the phase space. It is interesting to
see that this remains true even with the more intricate two-scale
kinematics that characterizes prompt photon production in the soft
limit, especially in view of the fact that the theoretical status of
Sudakov resummation for prompt photon production is rather less
satisfactory than for DIS or Drell-Yan. Also, this approach does not
require a separate treatment of individual colour structures when more
than one colour structure contributes to fixed order results.

The resummation formulae obtained here turn out
to be less predictive than previous results~\cite{dps,ctdp}: a higher
fixed-order computation is required in order to determine the
resummed result. This depends on the fact that here neither
specific factorization properties of the cross section in the soft
limit is assumed, nor that soft emission satisfies eikonal-like relations which
allow one to determine some of the resummation coefficients in terms
of universal properties of collinear radiation. Currently,  fixed-order results are only available
up to $O(\alpha_s^2)$ for prompt photon production. An order $\as^3$
computation is required to check nontrivial properties of the
structure of resummation: for example, factorization, whose effects only appear
at the next-to-leading log level, can only be tested at
$O(\alpha_s^3)$. The greater flexibility of the approach presented
here would turn out to be necessary if the prediction obtained
using the more restrictive resummation of refs.~\cite{dps,ctdp} were
to fail at order $\alpha_s^3$.

\appendix
\section{Appendix}
In this Appendix, we prove some properties of matrices built from
the coefficients
eq.~(\ref{Cijm}), which appear in the perturbative expansion of the
highest order contribution $\gamma_k$ to the resummed anomalous
dimension eq.~(\ref{gammagen},\ref{gammas}).

\begin{enumerate}
\item
The $k\times k$ matrix
\beq
A^{(k)}_{mi}=C^{(0,k-i)}_m;
\qquad 0\leq m \leq k-1,\qquad 0\leq i \leq k-1
\eeq
is non-singular.
\bigskip
\\
{\it Proof:} From eq.~(\ref{Cijm})
we see that $A^{(k)}_{mi}$ is a degree-$m$
polynomial in $i$:
\beq
A^{(k)}_{mi}=\frac{(-1)^{m+1}}{(m+1)!}
(m+k-i-1)\times\ldots\times(k-i)=\sum_{l=0}^m A_l\,i^l.
\eeq
It follows that a generic linear combination of the rows of $A^{(k)}$
\beq
\sum_{m=0}^{k-1}x_m\,A^{(k)}_{mi}
=\sum_{m=0}^{k-1}x_m\sum_{l=0}^m A_l\,i^l
=\sum_{l=0}^{k-1}A_l\,i^l\sum_{m=0}^l x_m
\eeq
can only vanish if $x_m=0$ for all $m$.

It follows that each term in the sum over $m$ in eq.~(\ref{DISDY})
provides a linearly independent condition on the coefficients
$g_{ik-i}$,
$0\le i\le k-1$.
\item
The $k\times k$ matrix
\beq
A^{(k)}_{mi}=C^{(k-i,0)}_m;
\qquad 0\leq m \leq k-1,\qquad 0\leq i \leq k-1
\eeq
is non-singular.
\bigskip
\\
{\it Proof:} This statement follows immediately from the previous one,
because it is easy to show that
\beq
\label{cond0}
C^{(k-i,0)}_m=2^m\,C^{(0,k-i)}_m.
\eeq

\item
\label{stat3}
The $(k+1)\times(k+1)$ matrix
\beq
B^{(k)}_{mj}=C^{(j,k-j)}_m;
\qquad 0\leq m \leq k,\qquad 0\leq j \leq k
\eeq
is non-singular. 
\bigskip
\\
{\it Proof:} This statement can be proved by induction on $k$.
For $k=1$ we have
\beq
B^{(1)}=\left(\begin{array}{cc}-1&-1\\1/2&1\end{array}\right)
\eeq
which is manifestly non-singular. We now assume that $B^{(k-1)}$
is non-singular, and we consider a linear combination of the columns of 
$B^{(k)}$:
\beq
\sum_{j=0}^kx_j\,B^{(k)}_{mj}=
(x_0+2^mx_k)\,C^{(0,k)}_m+\sum_{j=1}^{k-1}x_j\,C^{(j,k-j)}_m,
\eeq
where we have used eq.~(\ref{cond0}).
For $i,j\geq 1$ the following identity holds:
\beq
C_m^{(i,j)}=2C_m^{(i,j-1)}-C_m^{(i-1,j)}.
\label{cond1}
\eeq
Equation~(\ref{cond1}) can be verified directly, using the standard properties
of the binomial coefficients
\beq
\left(\begin{array}{c}n-1\\k-1\end{array}\right)
=
\left(\begin{array}{c}n\\k\end{array}\right)
-
\left(\begin{array}{c}n-1\\k\end{array}\right);\qquad
\left(\begin{array}{c}n\\k\end{array}\right)=0 \;{\rm for}\;n<k.
\eeq
Using eq.~(\ref{cond1}) we get
\bea
\sum_{j=0}^kx_j\,B^{(k)}_{mj}
&=&(x_0+2^mx_k)\,C^{(0,k)}_m
+2\sum_{j=1}^{k-1}x_j\,C^{(j,k-j-1)}_m
-\sum_{j=1}^{k-1}x_j\,C^{(j-1,k-j)}_m
\nonumber\\
&=&(x_0+2^mx_k)\,C^{(0,k)}_m
+2\sum_{j=1}^{k-1}x_j\,C^{(j,k-1-j)}_m
-\sum_{j=0}^{k-2}x_{j+1}\,C^{(j,k-1-j)}_m
\nonumber\\
&=&(x_0+2^mx_k)\,C^{(0,k)}_m
+\sum_{j=0}^{k-1}\bar x_j\,B^{(k-1)}_{mj},
\label{Bk}
\eea
where
\bea
&&\bar x_0=-x_1
\nonumber
\\
&&\bar x_j=2x_j-x_{j+1};\qquad 1\leq j\leq k-2
\label{xbar}
\\
&&\bar x_{k-1}=2x_{k-1}.
\nonumber
\eea
The linear combination in eq.~(\ref{Bk}) can only vanish if the
two terms are separately zero, since $C^{(0,k)}_m$ is a degree-$(k-1)$
polynomial in $m$, while $B^{(k-1)}_{mj}$ is at most of degree $k-2$.
Hence, for eq.~(\ref{Bk}) to vanish, it must be
\bea
&&x_0+2^m x_k=0
\\
&&\sum_{j=0}^{k-1}\bar x_j\,B^{(k-1)}_{mj}=0,
\eea
and therefore
\beq
x_0=0;\quad x_k=0;\quad \bar x_j=0
\eeq
by the induction hypothesis. Using eqs.~(\ref{xbar}) this gives
\beq
x_0=x_1=\ldots=x_k=0.
\eeq

It follows that eqs.~(\ref{condzero}-\ref{condkappa})
provides a linearly independent condition on the coefficients
$g_{0ik-i}$,
$0\le i\le k$.

\item 
\label{stat4}
Define an $M\times\frac{k(k+3)}{2}$ matrix $D^{(k)}$,
whose columns are the $M$-component vectors
\beq
D^{(k)}_m=C^{(j,k-i-j)}_m;\qquad
0\leq i\leq k-1;\quad 0\leq j\leq k-i;\qquad
0\leq m\leq M.
\eeq
The rank (number of linearly-independent columns) of $D^{(k)}$
is $2k$.
\bigskip
\\
{\it Proof:} We use induction on $k$. For $k=1$,
$D^{(1)}$ is a $2\times 2$ matrix with columns
\beq
D^{(1)}_m=\left(C_m^{(0,1)},\quad C_m^{(1,0)}\right)
=\frac{(-1)^{m+1}}{m+1}\,(1,\quad 2^m),
\eeq
that are linearly independent; the rank of $D^{(1)}$ is 2. Let us check
explicitly also the case $k=2$. In this case
g\beq
D^{(2)}_m=
\left(C_m^{(0,1)},\quad C_m^{(1,0)},
\quad C_m^{(0,2)},\quad C_m^{(1,1)},\quad C_m^{(2,0)}\right).
\eeq
The first two columns are the same as in the case $k=1$: 
they span a 2-dimensional subspace.
The last three columns are independent as a consequence of
statement~\ref{stat3} of this Appendix. Furthermore, $C_m^{(0,2)}$ and
$C_m^{(2,0)}=2^m C_m^{(0,2)}$ 
are independent of all
other columns, because they are the only ones that
are proportional to a degree-1 polynomial in $m$.
Finally, $C_m^{(1,1)}$ is a linear combination of the first two columns,
as a consequence of eq.~(\ref{cond1}) with
$i=j=1$. Thus, the rank of $D^{(2)}$ is $2+2=4$.

We now assume that $D^{(k-1)}$ has rank $2(k-1)$, and we write
the columns of $D^{(k)}$ as
\bea
&&D^{(k)}_m=
\left(C_m^{(j,k-1-i-j)},\quad C_m^{(l,k-l)}\right)
\\
&&0\leq i\leq k-2,\qquad 
0\leq j\leq k-1-i\qquad
\qquad 0\leq l\leq k.
\eea
By the induction hypothesis, only $2(k-1)$ of the columns  
$C_m^{(j,k-1-i-j)}$ are independent.
The columns $C_m^{(l,k-l)}$ are all independent as a consequence
of statement \ref{stat3}; among them,
those with $1\leq l\leq k-1$ can be expressed as
linear combinations of $C_m^{(j,k-1-i-j)}$ by eq.~(\ref{cond1}).
Only $C_m^{(0,k)}$ and $C_m^{(k,0)}$ are independent of
all other columns because they are proportional to a degree-$(k-1)$
polynomial in $m$, while all others are at most of degree $(k-2)$.
Hence, only two independent vectors are added to the 
$2(k-1)$-dimensional subspace spanned by $C_m^{(j,k-1-i-j)}$, and
the rank of $D^{(k)}$ is
\beq
2(k-1)+2=2k.
\eeq
It follows that each individual terms in the sum over $m$ in
eq.~(\ref{gammak}) depends only on $2k$ independent linear combinations of
the coefficients 
$g_{ijk-i-j},\> 0,\le i \le k-1,\>0\le j\le k-i$.

\end{enumerate}

\end{document}